\documentclass[reprint,final,floatfix,nofootinbib,superscriptaddress,onecolumn,eqsecnum,longbibliography]{revtex4-2}

\usepackage{lmodern,mathtools,amssymb,mathrsfs,bbm,microtype,bm} % fonts
\usepackage{color,graphicx} % figures
\graphicspath{{figures/}}
\definecolor{bluesmoke}{rgb}{0.207843,0.415686,0.623529}
\usepackage[
  allcolors=bluesmoke,
  colorlinks,
  pdftex,
  pdftitle={Optimizing diffusion-limited transport, with applications to electrochemical systems},
  pdfauthor={Manu Mannattil and L. Mahadevan},
  hypertexnames=false
]{hyperref}
\usepackage[textwidth=5.318in,textheight=8.594in,paperwidth=6.85in,paperheight=9.72in]{geometry}

\definecolor{hlcolor}{rgb}{0.646,0.165,0.165}

\DeclareMathOperator{\erfc}{erfc}
\DeclarePairedDelimiter{\abs}{\lvert}{\rvert}

\def\dd{\ensuremath\mathrm{d}}
\def\e{\ensuremath\mathrm{e}}
\def\unit{{\kern2.333pt}}

\def\Qc{\ensuremath{Q_{\text{c}}}} % capacity at constant current
\def\Qo{\ensuremath{Q^{\star}}} % optimal capacity
\def\TS{\ensuremath{T_{\text{S}}}} % Sand's time
\def\Tc{\ensuremath{T_{\text{c}}}} % constant (minimum) time
\def\To{\ensuremath{T^{\star}}} % optimal (minimum) time
\def\jc{\ensuremath{j_{\text{c}}}} % constant current
\def\jo{\ensuremath{j^{\star}}} % optimal current
\def\jr{\ensuremath{j_{\text{r}}}} % riding current
\def\ps{\ensuremath{p^{\star}}} % optimal multiplier
\def\ts{\ensuremath{t^{\star}}} % switching time
\def\us{\ensuremath{u^{\star}}} % optimal flux
\def\zc{\ensuremath{z}} % charge number

\def\unitj{\ensuremath{\unit\text{mA}\unit\text{cm}^{-2}}}
\def\unitt{\ensuremath{\unit\text{s}}}
\def\unitQ{\ensuremath{\unit\text{mAh}\unit\text{cm}^{-2}}}
\def\unitD{\ensuremath{\unit\text{cm}^{2}\unit\text{s}^{-1}}}

\def\jtheta{\ensuremath{\vartheta_{3}}}

\begin{document}

\title{Optimizing diffusion-limited transport, with applications to electrochemical systems}
\author{Manu Mannattil}
\affiliation{School of Engineering and Applied Sciences, Harvard University, Cambridge, Massachusetts 02138, USA}
\author{L.~Mahadevan}
\affiliation{School of Engineering and Applied Sciences, Harvard University, Cambridge, Massachusetts 02138, USA}
\affiliation{Department of Physics, Harvard University, Cambridge, Massachusetts 02138, USA}
\affiliation{Department of Organismic and Evolutionary Biology, Harvard University, Cambridge, Massachusetts 02138, USA}

\begin{abstract}
  \leftskip=0.65cm\rightskip=\leftskip
  Transport in certain systems fails when diffusion cannot replenish or remove material from a  boundary rapidly enough, causing the boundary concentration to reach a critical minimum or maximum.
  Motivated by experiments in electrochemical systems that demonstrate this diffusion-limited failure of charge transport, we determine dynamic current protocols that maximize charge transfer subject to a prescribed concentration constraint.
  The optimal protocol has a ``bang--ride'' structure: the maximum feasible current is used until the boundary concentration reaches its critical value, after which the current is progressively reduced to maintain that value.
  For semi-infinite domains, we derive the optimal currents analytically and obtain system-independent upper bounds on their improvement over constant-current operation.
  We also extend the framework to finite and multilayer domains and show that the same formulation applies to both charging and discharging.
\end{abstract}

\maketitle

% RSPA Subject Areas: applied mathematics, mathematical modelling, power and energy systems
% RSPA Keywords: diffusion, electrochemical systems

\section{Introduction}

Diffusion is among the most common forms of material transport, occurring in biological organisms to engineered devices~\cite{crank1979,cussler2009}.
In many such systems, when material is removed through a boundary faster than diffusion can replenish it, the boundary concentration may fall to zero, thereby providing a natural criterion for diffusion-limited failure.
Such limitations arise in plant roots where nutrient depletion at the root surface strains its uptake~\cite{willigen1994,willigen2017}.
Similarly, cell cultures can become severely oxygen-limited when cellular demand reaches the maximum diffusive supply, corresponding to near-zero oxygen concentrations at the cell surface~\cite{place2017}.
Bacterial growth can likewise become diffusion-limited when substrate depletion occurs at the cell surface~\cite{schulz2001}.
Related effects also arise during drying of porous materials, from clay bricks~\cite{pel2002} to fruits~\cite{varadharaju2001}, where rapid moisture loss at the outer surface can cause shrinkage before internal moisture equilibrates, leading to case hardening, wrinkling, or cracking~\cite{tsapis2005,gulati2015}.

Electrochemical systems such as batteries, supercapacitors, and fuel cells provide a particularly important setting for diffusion-limited transport~\cite{sand1901,bard2001,newman2021}.
During charge or discharge, an imposed current can consume material near an electrode faster than diffusion can replenish it, leading to performance degradation and, in severe cases, thermal runaway or device failure~\cite{bai2016,jana2019}.
Motivated by these observations, we ask: in a diffusion-limited electrochemical system, how should the imposed current vary in time to maximize material transport while preventing the boundary concentration from crossing a prescribed limit?

% Understanding charge--discharge dynamics and developing optimal (dis)charging protocols are important for improving the performance of electrochemical energy-storage systems such as batteries, supercapacitors, and fuel cells~\cite{gao2019,breitsprecher2018,hemi2015}.
% In such systems, one would like to choose the applied current as a function of time so as to maximize usable capacity while respecting transport and operating constraints.
% Direct experimental optimization can take months to years~\cite{severson2019,attia2020}, while machine-learning approaches can often obscure the underlying physics~\cite{attia2020,jiang2022}.
% Fully resolved multiphysics simulations are also costly and require many material parameters~\cite{brosa-planella2022,singla2026,singla2026a}.
% Consequently, simplified models, including equivalent-circuit models, have been widely used to optimize charging time, power loss, and temperature rise~\cite{hu2013,abdollahi2016,parvini2018}.

Several closely related problems have been studied previously.
Early work considered pulsed- and continuous-current protocols for diffusion-limited electrochemical transport~\cite{purushothaman2005,purushothaman2006,boovaragavan2007}.
More recently, such strategies have also been extended to nonlinear phase-field models~\cite{telmasre2024}.
In parallel, optimal-control studies based on single-particle battery models~\cite{brosa-planella2022} showed that the optimal fast-charging currents have a \emph{bang--ride} structure: the maximum allowable current is used until an operational constraint is reached, after which the current is reduced just enough to avoid violating the constraints~\cite{park2020}.
Subsequent work established broader necessary conditions for bang--ride controls~\cite{matschek2023,shi2025}.

The bang--ride structure of optimal currents found in previous studies is not, by itself, especially surprising.
It is closely related to complementary slackness in constrained optimization~\cite{boyd2004}: in such problems, the constraints define an admissible region in the space of possible solutions, and the optimal solution evolves freely in the interior but adjusts to follow the constraint boundary once it is reached.
Furthermore, previously reported performance gains have generally been obtained either for prescribed or numerically obtained protocols~\cite{purushothaman2005,purushothaman2006,boovaragavan2007} or using reduced-order and spatially discretized models~\cite{park2020,matschek2023,shi2025}, rather than by directly optimizing the underlying diffusion problem.
It therefore remains unclear how such gains relate to experimentally relevant limits such as Sand's capacity, which quantifies the amount of charge that can be transferred before diffusion limitation is reached, and constitutes an important design constraint in metal batteries~\cite{bai2016}.
In particular, it is not clear whether the resulting capacity gains are fundamentally bounded or, equivalently, how much charging or discharging time can ultimately be saved.

To address the aforementioned questions, in this paper, we first consider homogeneous linear diffusion in a semi-infinite domain, motivated by experiments on lithium (Li) electrodeposition~\cite{bai2016}.
Assuming that diffusion is the only limiting process, we derive analytically the bang--ride current protocol that allows the transferred charge to be maximized.
This yields a system-independent upper bound of $4/\pi - 1\approx 27\%$ on the capacity gain over constant-current protocols.
As a second example, we consider fast charging of Li-ion batteries~\cite{purushothaman2006}, where we show that bang--ride currents can reduce the charging time by up to $1-\pi^2/16\approx 38\%$ relative to constant-current charging.

Finally, we consider optimal stripping (discharging) protocols motivated by recent experiments on electrochemical cells with sodium--potassium (NaK) liquid anodes~\cite{wu2025}, where pulsed currents were shown to increase the usable capacity.
The experiments also revealed a current-dependent transition from stable to unstable stripping, attributed to sodium (Na) depletion at the anode--electrolyte interface~\cite{wu2025a}.
We model this behavior using a phenomenological two-layer anode comprising a transport-limiting interfacial region coupled to a higher-diffusivity reservoir.
Despite its simplicity, the model captures the experimental behavior well and again predicts an optimal bang--ride discharge protocol.

It should be emphasized that the optimal current profiles and performance bounds derived here account for diffusion alone.
In practical electrochemical systems, additional effects, including thermal constraints, power losses, reaction kinetics, and degradation mechanisms, must also be considered~\cite{arora1999,tomaszewska2019}.
Nevertheless, diffusion often constitutes a dominant limitation during (dis)charging, so our results provide a useful fundamental benchmark for the gains achievable through current control.

This paper is organized as follows.
Section~\ref{sec:electrodeposition} examines protocols for maximizing charge deposition in Li electrodeposition experiments and introduces the general bang--ride structure for diffusion-limited transport.
Appendices~\ref{app:optimality} and~\ref{app:transfer} provide detailed derivations of the optimal protocol and the diffusion transfer function.
We then consider fast Li-ion battery charging in Sec.~\ref{sec:charging} and optimal stripping of NaK anodes in Sec.~\ref{sec:nak}, before concluding in Sec.~\ref{sec:conclusion}.

\begin{figure}{\centering\includegraphics[scale=1.0]{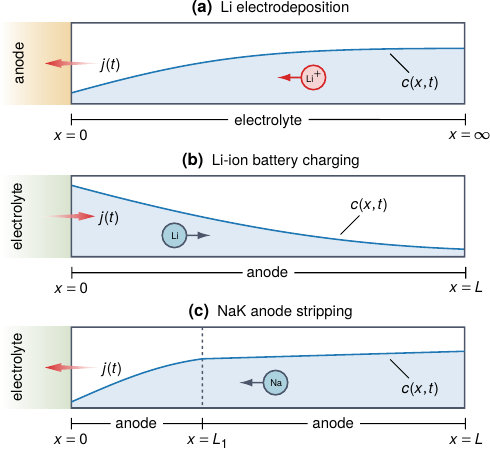}}
  \caption{Schematics of the diffusion problems considered in this paper.
    Each panel shows the domain, the direction of material transfer due to the applied current density $j(t)$, a typical concentration profile $c(x,t)$, and the diffusing species.
    The three cases are (a) Li electrodeposition (Sec.~\ref{sec:electrodeposition}), (b) Li-ion battery charging (Sec.~\ref{sec:charging}), and (c) NaK anode stripping (Sec.~\ref{sec:nak}).
  }
  \label{fig:diffusion}
\end{figure}

\section{Optimal Lithium electrodeposition}
\label{sec:electrodeposition}

In Li-ion batteries, charging drives Li$^{+}$ ions from the bulk electrolyte toward the anode--electrolyte interface, where they are reduced and incorporated into the anode either by intercalation or by electrodeposition as metallic Li.
At sufficiently high charging currents, diffusion may be unable to replenish Li$^{+}$ ions near the interface as rapidly as it is consumed.
For currents that exceed this diffusion limitation, the Li$^{+}$ concentration at the anode surface can approach zero, leading to nonuniform deposition and dendritic Li growth.
Such dendrites can cause internal short circuits and thermal runaway and, in severe cases, lead to explosions~\cite{tarascon2001,jana2019}.
Experiments on Li electrodeposition designed to mimic battery charging have observed precisely this behavior, with a transition from relatively benign ``mossy'' deposition to hazardous dendritic growth~\cite{bai2016}.

\subsection{Diffusion in semi-infinite domains}

To understand Li electrodeposition, we begin with a generic one-dimensional (1D) diffusion problem on the semi-infinite half-line, $x \geq 0$.
We identify this domain with the electrolyte adjacent to the electrode onto which Li is deposited, with $x=0$ representing the anode--electrolyte interface, and $x>0$ extending into the bulk electrolyte as shown in Fig.~\ref{fig:diffusion}(a).
The semi-infinite approximation is appropriate when the characteristic length scale over which the concentration varies remains small compared with the electrode separation, as is typically the case in Li batteries.

Let $c(x,t)$ denote the concentration of Li$^{+}$ ions in the electrolyte, which evolves according to an effective diffusion equation~\cite{newman2021}
\begin{equation}
  \partial_{t}c(x,t) = D\partial_{x}^{2}c(x,t),
  \quad
  x \geq 0,
  \enspace
  t \geq 0,
  \label{eq:inf_diffusion}
\end{equation}
where $D$ is the diffusion coefficient.
We assume a spatially uniform initial concentration, $c(x,0)=c_{0}$.
During electrodeposition, Li$^{+}$ ions are depleted at the boundary $x=0$, where they are reduced to neutral Li atoms because of the applied current.
The resulting interfacial Faradaic flux $u(t)$ is related to current density $j(t)$ through~\cite{bard2001}
\begin{equation}
  D\partial_{x}c(0,t) = u(t)
  = \left(\frac{\tau_{\text{a}}}{\zc F}\right)j(t),
  \label{eq:inf_bc}
\end{equation}
where $\tau_{\text{a}}$ is the anionic transference number, representing the fraction of the electrical current carried by the negative electrolyte ions, $\zc$ is the charge number of Li$^{+}$, and $F$ is the Faraday constant~\cite{bard2001}.

For a time-dependent interfacial flux $u(t)$, Duhamel's principle~\cite{stone2009} gives the solution to the diffusion equation, Eq.~\eqref{eq:inf_diffusion}, subject to the boundary condition in Eq.~\eqref{eq:inf_bc}, as
\begin{equation}
  c(x, t) = c_{0} + u(0)w(x, t) + \int_{0}^{t}\dd{s}\, \frac{\dd u(s)}{\dd s}\, w(x, t - s).
  \label{eq:inf_duhamel}
\end{equation}
Here, $w(x,t)$ is the solution to the diffusion equation $\partial_t w(x,t)=D\partial_x^2 w(x,t)$ on the half-line, with the initial condition $w(x,0)=0$ and the boundary condition $D w_x(0,t)=1$.
Using the method of images~\cite{pinsky2011} and the particular solution $w = x/D$, we find
\begin{equation}
  w(x, t) = \frac{1}{D}
  \left[
  x\erfc\left(\frac{x}{\sqrt{4 D t}}\right)
  - \sqrt{\frac{4 D t}{\pi}}\exp\left(-\frac{x^{2}}{4 D t}\right)
  \right],
  \label{eq:sands_w}
\end{equation}
where $\erfc(\cdot)$ is the complementary error function. Substituting $w(x, t)$ in Eq.~\eqref{eq:inf_duhamel} and integrating by parts once, we find the solution to Eq.~\eqref{eq:inf_diffusion} in terms of the interfacial flux $u(x)$ as
\begin{equation}
  c(x, t) = c_{0} - \frac{1}{\sqrt{\pi D}}\int_{0}^{t}\dd{s}\,\frac{u(s)}{\sqrt{t-s}}\exp\left[-\frac{x^{2}}{4D(t-s)}\right].
  \label{eq:semi_infinite}
\end{equation}

From the full solution, Eq.~\eqref{eq:semi_infinite}, we find the boundary concentration to be
\begin{equation}
  c(0, t) = c_{0} - \int_{0}^{t}\dd{s}\,Z(t-s)\,u(s),
  \quad
  \text{with}
  \quad
  Z(t) = \frac{1}{\sqrt{\pi Dt}}.
  \label{eq:inf_transfer}
\end{equation}
This equation generalizes the classical Sand's equation~\cite{sand1901,bard2001} to a time-varying flux $u(t)$, or equivalently, a time-varying current density $j(t)$, with the two related by Eq.~\eqref{eq:inf_bc}.
As the transfer function $Z(t)$ is positive and strictly decreasing, a nonzero Faradaic flux causes the boundary concentration $c(0,t)$ to decrease with time.

For a constant Faradaic flux $u_{\text{c}}$ introduced by a constant current $\jc$, Sand's time $\TS$ is the time at which the boundary concentration $c(0, t)$ first vanishes.
Using Eq.~\eqref{eq:inf_bc} and Eq.~\eqref{eq:inf_transfer}, we find
\begin{equation}
  \TS = \pi D\left(\frac{c_{0}}{2u_{\text{c}}}\right)^{2} = \pi D \left(\frac{c_{0}\zc F}{2\tau_{\text{a}}\jc}\right)^{2}.
  \label{eq:inf_sands_time}
\end{equation}
In Li electrodeposition experiments, Sand's time has been considered as a physically meaningful estimate of when the electrolyte concentration at the electrode surface is depleted under constant current and indicates the onset of dendrite growth~\cite{bai2016}.

\subsection{Dynamic currents for optimal electrodeposition}

We now turn to the problem of finding the current profile that maximizes the areal capacity $Q$, which is the charge transferred per unit area across the anode--electrolyte interface, over a prescribed time $T$. To avoid diffusion limitation, we require that the boundary concentration $c(0,t)$ remain above a fraction $\chi$ of the initial concentration $c_{0}$.
Additionally, we assume that the current density satisfies $0 \leq j(t) \leq j_{0}$, where $j_{0}$ is the maximum allowable current density.
Thus, the optimization problem for electrodeposition takes the form
\begin{gather}
  \label{eq:objective}
  \underset{j(t)}{\text{max}}\quad Q = \int_{0}^{T}\dd{t}\,j(t);\\
  \text{subject to}\quad
  \left\{
  \begin{aligned}
    c(0,t)      & \geq \chi c_{0}, \\
    \enspace
    0 \leq j(t) & \leq j_{0}.
  \end{aligned}
  \right.
\end{gather}

Because the objective $Q$ increases monotonically with the current density $j$, the optimal protocol uses the maximum admissible current density $j_{0}$ for as long as the boundary-concentration constraint is satisfied.
Let $\ts$ denote the time at which the boundary concentration first reaches its allowable minimum, i.e., $c(0,\ts)=\chi c_{0}$.
Beyond $\ts$, maintaining $j=j_{0}$ would violate the constraint, while reducing the current more than necessary would decrease the objective.
The natural choice for $t\geq\ts$ is therefore the current density that maintains $c(0,t)=\chi c_{0}$, which we refer to as the riding current density $\jr(t)$.
Assuming that $\jr(t)$ exists and remains admissible, i.e., $0 \leq \jr(t) \leq j_0$, the optimal \emph{bang--ride} current density for the optimization problem in Eq.~\eqref{eq:objective} is
\begin{equation}
  \jo(t)=
  \begin{cases}
    j_{0},  & t < \ts;    \\
    \jr(t), & t \geq \ts,
  \end{cases}
  \label{eq:bang_ride}
\end{equation}
where the switching time $\ts$ is defined as the solution to $c(0, \ts) = \chi c_{0}$ in the bang phase.

The bang--ride structure of the optimal current is intuitive and, for diffusion-limited systems, can be established explicitly as shown in Appendix~\ref{app:optimality}.
Related bang--ride optimality results have also been obtained for fast-charging problems based on reduced-order, spatially discretized single-particle battery models~\cite{park2020,matschek2023,shi2025}.
In contrast, the transfer-function formulation in Eq.~\eqref{eq:inf_transfer} preserves the full diffusion dynamics, allowing the currents to be computed exactly.

To find the optimal current, we only need to determine the switching time $\ts$ and the riding current density $\jr(t)$, since the current density during the bang phase is fixed at $j_{0}$.
In principle, $\jr(t)$ can be obtained directly from Eq.~\eqref{eq:inf_transfer} by solving the Volterra integral equation subject to the boundary-concentration constraint $c(0,t)=\chi c_{0}$ for $t\geq\ts$.
In practice, however, it is more convenient to determine $\jr(t)$ by directly solving the diffusion equation, Eq.~\eqref{eq:inf_diffusion}, as we do below.

\begin{figure}{\centering\includegraphics{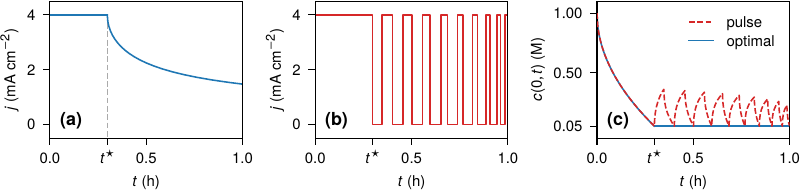}}
  \caption{(a) The current density in Eq.~\eqref{eq:inf_optimal}, for optimal Li electrodeposition using parameters relevant to the experiments in Ref.~\cite{bai2016}.
    The initial electrolyte concentration is $c_{0}=1\unit\text{M}$, the total deposition time is $T=1\unit\text{h}$, and the maximum current density is $j_{0} = 4\unitj$.
    The current initially takes the maximum (bang) value $j_{0}$ and switches to the riding current $\jr(t)$ at $\ts \approx 0.3\unit\text{h}$.
    (b) Pulse-width-modulated approximation of the optimal current density in (a) using 10 pulses.
    (c) Boundary concentration $c(0,t)$ for the optimal current (solid) and its pulsed approximation (dashed).
    In both cases, the concentration stays above a fraction $\chi = 0.05$ of its initial value.
  }
  \label{fig:li_pulse}
\end{figure}

In terms of the Faradaic flux $u_{0}$ corresponding to the maximum (bang) current density and using Eq.~\eqref{eq:inf_transfer}, the switching time for semi-infinite diffusion is
\begin{equation}
  \ts = \pi D\left[(1 - \chi)\frac{c_{0}}{2u_{0}}\right]^{2} =
  \pi D \left[\frac{(1-\chi)c_{0}\zc F}{2\tau_{\text{a}}j_{0}}\right]^{2}.
  \label{eq:inf_switch}
\end{equation}
Meanwhile, from Eq.~\eqref{eq:inf_duhamel}, the concentration profile at the switching time is,
\begin{equation}
  c(x, \ts) = c_{0} - \frac{u_{0}}{D}
  \left[\sqrt{\frac{4 D t}{\pi}}\exp\left(-\frac{x^{2}}{4 D \ts}\right) - x\erfc\left(\frac{x}{\sqrt{4 D \ts}}\right)\right],
\end{equation}
To determine the riding current density for $t\geq\ts$, we then solve the following diffusion equation on the semi-infinite half-line:
\begin{equation}
  \left.
  \begin{aligned}
    \partial_{t}g(x, t) & = D \partial_{x}^{2}g(x, t),\\
    g(x, \ts)           & = c(x, \ts),                \\
    g(0, t)             & = \chi c_{0},
  \end{aligned}
  \right\}
  \quad
  x \geq 0,
  \enspace
  t \geq \ts.
\end{equation}
Using the standard diffusion kernel $W(x,t)=(4\pi Dt)^{-1/2}\exp[-x^{2}/(4Dt)]$ and the method of images, the solution to the above equation can be formally written as~\cite{carslaw1959}
\begin{equation}
  g(x, t) = \chi c_{0}
  + \int_{0}^{\infty}\dd{\xi}\,
  \left[W(x-\xi,t-\ts)-W(x+\xi,t-\ts)\right]
  \left[c(\xi,\ts)-\chi c_{0}\right].
\end{equation}
The riding current density is then obtained from the interfacial Faradaic flux $D\partial_{x}g(0, t)$ computed from the above solution.

For semi-infinite diffusion, the optimal current density takes the final form
\begin{equation}
  j(t) =
  \begin{dcases}
    j_{0},                                           & t < \ts;    \\
    \frac{2j_{0}}{\pi}\sin^{-1}\sqrt{\frac{\ts}{t}}, & t \geq \ts,
  \end{dcases}
  \label{eq:inf_optimal}
\end{equation}
where the switching $\ts$ is given by Eq.~\eqref{eq:inf_switch}.
Note that $\jr(t)$ is independent of the total electrodeposition time $T$.
Although a semi-infinite riding current permits charge transfer indefinitely as $t\to\infty$, any finite physical domain is eventually depleted.
We assume that this does not occur within the time interval $T$.

For the Li electrodeposition experiments of Ref.~\cite{bai2016}, the diffusivity $D=10^{-6}\unitD$, anionic transference number $\tau_{\text{a}} = 0.62$, and charge number $\zc=1$~\cite{bai2016}.
We also choose the initial Li$^{+}$ concentration to be $c_{0} = 1\unit\text{M}$.
The optimal current profile obtained using these parameters is shown in Fig.~\ref{fig:li_pulse}(a).
The current reaches the concentration constraint at $\ts\approx0.3\unit\text{h}$ and subsequently follows the riding profile.
Figure~\ref{fig:li_pulse}(b) shows a pulse-width-modulated approximation to the optimal current using 10 pulses.
Such approximations are not unique and depend on the number of pulses.
We require the boundary constraint $c(0, t) = \chi c_{0}$ to be satisfied at the end of each ``on'' pulse and optimize the intervening ``off'' times to maximize the objective in Eq.~\eqref{eq:objective}.
This yields a nonlinear optimization problem for the pulse widths, with the optimal ``on'' durations decreasing over time.
Figure~\ref{fig:li_pulse}(c) compares the corresponding boundary concentrations $c(0,t)$.
Under the continuous optimal current, $c(0,t)$ remains at the limiting concentration after the switching time $\ts$, whereas for pulsed currents it partially recovers during each ``off'' period before decreasing again.
The pulsed protocol therefore transfers less material than the continuous optimal current over the same time interval.

\begin{figure}{\centering\includegraphics{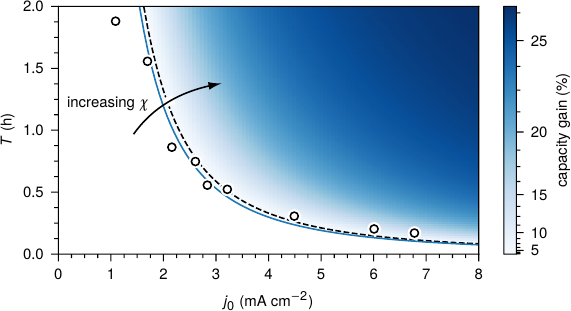}}
  \caption{Phase diagram in the $(j_{0},T)$ plane showing the conditions under which optimal currents increase the total areal capacity beyond constant currents for Li electrodeposition, using parameters relevant to the experiments in Ref.~\cite{bai2016}.
    Here, $j_{0}$ denotes the maximum (bang) current density and $T$ the total electrodeposition time.
    No capacity increase occurs unless $T$ exceeds the switching time $\ts$ given in Eq.~\eqref{eq:inf_switch}.
    For large $j_{0}$, the capacity gain approaches its maximum value of 27\%, as indicated by the shaded region.
    Sand's time $\TS$, Eq.~\eqref{eq:inf_sands_time}, computed using a constant current density equal to $j_{0}$, is indicated by the dashed curves.
    Optimal currents ensure safe electrodeposition even even when $T \gg \TS$ by keeping the boundary concentration above a fraction $\chi = 0.05$ of its initial value.
    The capacity-gain phase boundary shifts in the indicated direction as $\chi$ increases.
    The open circles show the experimental $\TS$ values from Ref.~\cite{bai2016}.
  }
  \label{fig:sands}
\end{figure}

\subsection{Capacity gains over constant-current electrodeposition}

Although the optimal current is time dependent, it is useful to compare its performance with the best admissible constant current over the same time horizon $T$.
Assuming no upper bound on the applied current, let $\jc$ be the largest constant current density for which $c(0,t)\geq \chi c_0$.
Since the boundary concentration decreases monotonically under constant current, $\jc$ must satisfy $c(0,T)=\chi c_0$.
Then, from Eq.~\eqref{eq:inf_transfer}, we find
\begin{equation}
  \jc^{2} =
  \frac{\pi D}{T}
  \left[\frac{(1-\chi)c_0 \zc F}{2\tau_{\text{a}}}\right]^{2}
  \quad
  \text{and}
  \quad
  \Qc=\jc T = \sqrt{\pi DT} \left[\frac{(1-\chi)c_0 \zc F}{2\tau_{\text{a}}}\right].
  \label{eq:Qc}
\end{equation}
The areal capacity at constant current $\Qc$ is closely related to, but is distinct from, the so-called Sand's capacity, which is the maximum charge per unit area that can be transferred at a given constant current without reaching diffusion limitation~\cite{bai2016,jana2019}.
In contrast, $\Qc$ denotes the maximum areal capacity achievable in a prescribed time interval $T$ using the largest constant current that remains below the diffusion limit.
For the optimization problem in Eq.~\eqref{eq:objective}, where the time interval $T$ is fixed, $\Qc$ provides a more appropriate benchmark than Sand's capacity for comparison with the optimal solution.

The charge transferred per unit area by the optimal current density in Eq.~\eqref{eq:inf_optimal} is
\begin{equation}
  \Qo = \int_0^T\dd{t}\,j(t) = j_0T\, \gamma\left(\frac{\ts}{T}\right), \quad T\geq \ts,
  \label{eq:inf_Q}
\end{equation}
where the function
\begin{equation}
  \gamma(s)
  =
  \frac{2}{\pi}
  \left[\sin^{-1}(\sqrt{s}) +\sqrt{s(1-s)}\right],
  \quad 0 \leq s \leq 1.
  \label{eq:inf_gamma}
\end{equation}
Clearly, $\gamma$ is bounded and monotonically increasing, with $0\leq\gamma(s)\leq1$.
As the allowable maximum current density $j_0\to\infty$, the switching time $\ts\to0$ and $\gamma(\ts/T)\to0$.
Nevertheless, the product $j_{0}\gamma(\ts/T)$ in Eq.~\eqref{eq:inf_Q} approaches a finite limit and we find
\begin{equation}
  \lim_{j_0\to\infty}\Qo
  =
  \frac{4}{\pi}
  \sqrt{\pi DT}
  \left[\frac{(1-\chi)c_0 \zc F}{2\tau_{\text{a}}}\right]
  =
  \frac{4}{\pi}\Qc
  \approx
  1.27\,\Qc.
  \label{eq:inf_Qo}
\end{equation}
Therefore, remarkably, even in the absence of an upper bound on possible currents, optimal currents can increase the transferable charge by about $4/\pi -1 \approx 27\%$ compared to the best constant-current protocol.
This parameter-independent upper bound is the maximum improvement achievable by optimal currents over constant currents in homogeneous semi-infinite diffusion.

For Li electrodeposition experiments, this bound limits the capacity gains achievable solely by controlling the applied current.
This limitation is illustrated more clearly by the phase diagram in Fig.~\ref{fig:sands}, shown in the $(j_{0},T)$ plane, which identifies the regimes in which optimized current profiles increase the areal capacity.
Capacity gains occur only when the total deposition time satisfies $T>\ts$, so that the protocol enters the riding regime.
Once this occurs, the optimal protocol always outperforms constant-current operation.
The shaded region shows the percentage capacity gain and indicates that the upper bound of 27\% can be approached even at moderate currents and deposition times.

The semi-infinite electrodeposition problem considered thus far provides both an explicit optimal protocol and a parameter-independent benchmark for the gains achievable through current control.
Although framed as electrodeposition, the problem is governed by depletion of Li$^{+}$ in the electrolyte near the anode--electrode boundary.
We next consider the reverse transport process, in which a species accumulates instead of depleting at the boundary.
This leads naturally to the problem of fast charging a finite anode.

\section{Fast charging of lithium-ion batteries}
\label{sec:charging}

Typical Li-ion batteries use graphite anodes.
During charging, Li$^+$ ions are reduced at the anode--electrolyte interface to neutral Li atoms, which intercalate into the graphite and diffuse through the anode.
Because the anode can accommodate only a finite Li concentration, large charging currents can cause Li to accumulate at the interface faster than it can be transported into the graphite~\cite{jia2026}.
If metallic Li precipitates at the interface, it may result in capacity losses and is a potential safety hazard~\cite{arora1999}.
This motivates the problem of selecting the current to minimize the charging time while ensuring that the interfacial Li concentration remains below a critical, saturation concentration throughout the charging process~\cite{purushothaman2006}.

\subsection{Finite-domain diffusion}

The diffusion problem considered in the previous sections also describes Li insertion into a solid anode, with the direction of transport reversed.
Unlike the semi-infinite setting considered previously, we now account for the finite anode size by assuming that it occupies a finite region of length $L$.
Assuming that the diffusivity of intercalated Li is $D$, the Li concentration $c(x,t)$ satisfies the diffusion equation
\begin{equation}
  \partial_{t}c(x, t) = D\partial_{x}^{2}c(x, t),
  \quad
  0 \leq x \leq L,
  \enspace
  t \geq 0.
  \label{eq:finite_diffusion}
\end{equation}
The anode is assumed to be initially unlithiated with an initial Li concentration $c(x, 0) = 0$.
As the battery charges, Li crosses the anode--electrolyte interface at $x=0$ and proceeds to diffuse inward through the anode [see Fig.~\ref{fig:diffusion}(b)].
Thus, compared to the Li$^{+}$ depletion problem considered previously, here, the interfacial flux is reversed.
Taking the signs of both the applied current density $j(t)$ and the resulting Faradaic flux $u(t)$ at the interface to be positive, the boundary conditions are~\cite{bard2001}
\begin{equation}
  -D\partial_{x}c(0, t) = u(t) = \left(\frac{1}{\zc F}\right)j(t)
  \quad \text{and} \quad
  \partial_{x}c(L, t) = 0.
  \label{eq:finite_bc}
\end{equation}

The finite-domain diffusion equation defined by Eqs.~\eqref{eq:finite_diffusion} and \eqref{eq:finite_bc} also admits an exact transfer-function solution for the boundary concentration, which takes the form (see Appendix~\ref{app:transfer})
\begin{equation}
  c(0,t) = \int_{0}^{t}\dd{s}\,Z(t - s)\,j(s),
  \quad
  \text{with}
  \quad
  Z(t)
  =
  L^{-1}\jtheta\left(0, \e^{-\pi^{2}Dt/L^{2}}\right).
  \label{eq:finite_transfer}
\end{equation}
Here, $\jtheta$ is the Jacobi theta function~\cite{olver2010} $\jtheta(z, q) = 1 + 2\sum_{m=1}^{\infty} q^{m^{2}} \cos(2mz)$.
As with the semi-infinite case, the transfer function $Z(t)$ here is strictly decreasing and positive.
Using the series expansion $\jtheta(0, \e^{-q}) = \sqrt{\pi/q}[1 - q^{2}/48 + \mathcal{O}(q^{3})]$ for $q \to 0$, we see that, when $L \to \infty$, Eq.~\eqref{eq:finite_transfer} reduces to the semi-infinite boundary concentration in Eq.~\eqref{eq:semi_infinite}.

\subsection{Current profiles for fast charging}

The goal of a fast-charging protocol is to minimize the total time required to transfer a prescribed charge per unit area $Q_{0}$, while respecting the boundary-concentration constraint $c(0,t)\leq c_{0}$ and the current-density constraint $0\leq j(t)\leq j_{0}$.
Thus, the optimization problem reads
\begin{gather}
  \underset{j(t),\,T}{\text{min}}\quad T = \int_{0}^{T}\dd{t};\\
  \text{subject to}\quad
  \left\{
  \begin{aligned}
    \enspace\int_{0}^{T}\dd{t}\,j(t) & = Q_{0},    \\
    c(0,t)                           & \leq c_{0}, \\
    0 \leq j(t)                      & \leq j_{0}.
  \end{aligned}
  \right.
\end{gather}

We now show that the fast-charging optimal current also takes a bang--ride form.
To do so, first consider the related problem of maximizing the total charge transferred over a fixed time $T$, subject to the same constraints $c(0,t)\leq c_{0}$ and $0\leq j(t)\leq j_{0}$.
From the previous section, the optimal current density $\jo(t)$ for this fixed-time problem has a bang--ride form.
Denote by $\Qo(T)$ the maximum charge per unit area that can be transferred in time $T$ using this protocol.

The minimum transfer time is then the first time $\To$ at which the maximum transferable charge reaches the prescribed value $Q_{0}$, i.e.,
\begin{equation}
  \Qo(\To) = Q_{0}.
\end{equation}
To see this, suppose instead that there exists another feasible protocol $\tilde{j}(t)$ that transfers $Q_{0}$ in a shorter time $\widetilde{T}<\To$.
Since $\Qo(\widetilde{T})$ is, by definition, the maximum charge that can be transferred by any feasible protocol in the time interval $[0,\widetilde{T}]$, the existence of $\tilde{j}(t)$ would imply
\begin{equation}
  \Qo(\widetilde{T})
  \geq
  \int_{0}^{\widetilde{T}}\dd{t}\,\tilde{j}(t)
  =
  Q_{0}.
\end{equation}
This contradicts the definition of $\To$ as the first time at which $\Qo(T)$ reaches $Q_{0}$.
Therefore, no feasible protocol can transfer the prescribed charge in a time shorter than $\To$.
Since the fixed-time charge-maximizing protocol $\jo(t)$ transfers exactly $Q_{0}$ at $T=\To$, it is also the minimum-time charging protocol.
Hence, the optimal fast-charging current retains the same bang--ride form of Eq.~\eqref{eq:inf_optimal}.

For fast charging, it is more convenient to compute the optimal current numerically, instead of solving the Volterra integral equation $c(0,t)=c_{0}$ that follows from Eq.~\eqref{eq:finite_transfer}.
We first solve the diffusion equation, Eq.~\eqref{eq:finite_diffusion}, using a Chebyshev spectral method~\cite{trefethen2000}, subject to the boundary conditions in Eq.~\eqref{eq:finite_bc}, up to the switching time $\ts$.
The resulting concentration profile $c(x,\ts)$ is then used as the initial condition for the subsequent diffusion problem with mixed boundary conditions $c(0,t)=c_{0}$ and $\partial_{x}c(L,t)=0$.
For $t\geq\ts$, the riding current density is obtained from the boundary flux as $\jr(t)= FD\partial_{x}c(0,t)$.

\begin{figure}{\centering\includegraphics{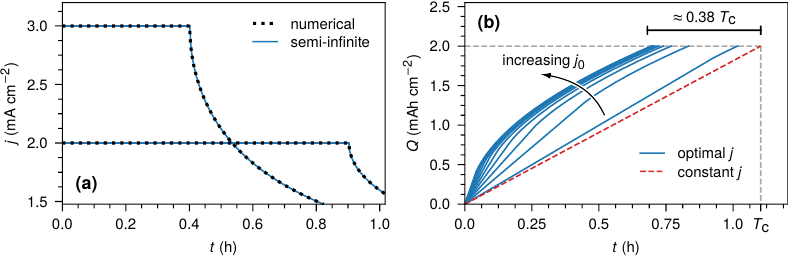}}
  \caption{(a) Optimal bang--ride current-density profiles for charging an anode to an areal capacity of $2\unitQ$  using parameters representative of a commercial Sony US18650S Li-ion battery~\cite{purushothaman2006}.
    Results are shown for maximum allowable current densities $j_{0}=2\unitj$ and $j_{0}=3\unitj$, comparing the numerical solutions with the semi-infinite approximations.
    (b) Areal capacity $Q$ as a function of time during charging under the optimal protocols (blue curves) and the best constant-current protocol satisfying the boundary-concentration constraint (dashed red curve).
    The constant-current protocol uses a current density of $\jc \approx 1.8\unitj$ and takes $T_{\text{c}} \approx 1.1\unit\text{h}$ to charge the anode to its maximum capacity.
    Optimal protocols reduce this time by up to $38\%$ as the maximum allowable current density $j_{0}$ increases.
    In the panel, $j_{0}$ is varied in 10 equal increments from $2\unitj$ to $10\unitj$.
  }
  \label{fig:charging}
\end{figure}

To illustrate the optimal fast-charging protocol, we use parameters representative of a commercial Li-ion battery~\cite{purushothaman2006}.
We take $c_{0}=77.0\unit\text{M}$ as the saturation concentration, corresponding approximately to the maximum theoretical molar concentration of pure solid Li~\cite{rumble2024}.
We use an anode thickness $L=128\unit\text{\textmu m}$, a Li diffusivity $D=3\times10^{-10}\unitD$, and charge number $\zc = 1$~\cite{purushothaman2006}.
Figure~\ref{fig:charging}(a) shows the numerically obtained optimal current profiles for two values of the maximum current density $j_{0}$.
Both protocols deliver the same total charge, but the charging time decreases as the maximum current density increases, as expected for bang--ride charging.
Figure~\ref{fig:charging}(a) also shows the optimal current profiles obtained under the semi-infinite approximation.
These profiles have the same form as Eq.~\eqref{eq:inf_optimal}, derived previously, and agree closely with the numerical solutions for the finite anode.

\subsection{Charging-time gains over constant currents}

Because the semi-infinite approximation closely reproduces the finite-domain optimal current profiles, we can use it to determine the fastest constant-current protocol that transfers the same charge per unit area, $Q_{0}$.
Let $\jc$ denote this constant current density and $\Tc$ the corresponding charging time.
For a constant-current protocol, the fastest admissible current is the one for which the boundary concentration reaches the limiting value $c_{0}$ exactly at $t=\Tc$.
Using Eq.~\eqref{eq:inf_transfer}, we see that $\jc$ and $\Tc$ satisfy a relation analogous to Sand's time, Eq.~\eqref{eq:inf_sands_time}, namely $\Tc=\pi D(c_{0}\zc F/2\jc)^{2}$.
Combining this relation with the charge constraint $\jc\Tc=Q_{0}$ gives the optimal constant current and its associated charging time,
\begin{equation}
  \jc = \frac{\pi D(c_{0} \zc F)^{2}}{4Q_{0}}
  \quad
  \text{and}
  \quad
  \Tc = \frac{1}{\pi D}\left(\frac{2Q_{0}}{c_{0}\zc F}\right)^{2}.
\end{equation}

For the semi-infinite optimal bang--ride protocol in Eq.~\eqref{eq:inf_optimal}, the transferred charge $Q_{0}$ and charging time $\To$ are instead given by Eq.~\eqref{eq:inf_Q}.
This yields,
\begin{equation}
  Q_{0} = j_{0}\To\gamma\left(\frac{\ts}{\To}\right),
\end{equation}
where $\ts = \pi D(c_{0}\zc F/2j_{0})^{2}$ is the semi-infinite switching time and the function $\gamma$ is defined in Eq.~\eqref{eq:inf_gamma}.

The minimum possible charging time under optimal charging is obtained in the limit $j_{0}\to\infty$, corresponding to an arbitrarily large current during the initial bang phase.
Using Eqs.~\eqref{eq:inf_gamma} and \eqref{eq:inf_Qo}, and the corresponding large-$j_{0}$ limit of the above expression for $Q_{0}$, we find
\begin{equation}
  \To = \frac{\pi^{2}}{16}\left[\frac{1}{\pi D}\left(\frac{2Q_{0}}{c_{0}\zc F}\right)^{2}\right] = \frac{\pi^{2}}{16}\Tc \approx 0.62\,\Tc.
\end{equation}
Thus, within the semi-infinite approximation, optimal bang--ride charging can reduce the charging time by up to approximately $1-\pi^2/16 \approx 38\%$ relative to the fastest admissible constant-current protocol.
As shown in Fig.~\ref{fig:charging}(b), the charging time decreases as the bang current density $j_{0}$ increases, approaching this $38\%$ reduction in the large-$j_{0}$ limit.

The preceding examples show that the optimal bang--ride structure survives a change from depletion to accumulation and from semi-infinite to finite diffusion, in a homogeneous transport medium. We now turn to a system in which spatially heterogeneous transport is essential, leading to consideration of a multi-layer diffusion problem.

\section{Optimal stripping of sodium--potassium anodes}
\label{sec:nak}

Na batteries have attracted increasing attention as lower-cost, resource-abundant alternatives to Li-ion batteries, particularly for large-scale energy storage~\cite{delmas2018}.
In this context, a recent study~\cite{wu2025} examined electrochemical cells comprising a NaK alloy anode in contact with a Na superionic conductor ($\text{Na}_{3}\text{Zr}_{2}\text{Si}_{2}\text{PO}_{12}$, NASICON) solid electrolyte.
In these cells, the NaK anode supplies Na, which is oxidized to Na$^{+}$ ions at the anode-electrolyte interface and is subsequently transported through the electrolyte.
Like other electrochemical systems, the attainable areal capacity of the NaK anode decreases with increasing stripping (discharge) currents because Na diffusion from the bulk cannot replenish the anode-electrolyte interface rapidly enough.
In addition, the charge-transfer resistance for Na oxidation increases as the interfacial Na concentration decreases~\cite{bard2001}.

The key experimental findings of Ref.~\cite{wu2025} can be summarized as follows.
At current densities below $0.6\unitj$, stripping was found to be stable, although still limited by Na diffusion.
Beyond this threshold, stripping became highly unstable.
For example, at a constant stripping current density of $0.25\unitj$, the cell remained stable up to a capacity of $20\unitQ$, whereas increasing the current density to $1\unitj$ reduced the attainable capacity to less than $1\unitQ$.
To mitigate the transport limitations caused by slow diffusion, a pulsed stripping protocol was used with a fixed current-on period of $t_{\text{on}} = 1\unitt$ and current-off periods $t_{\text{off}}$ ranging from $0.1\unitt$ to $10\unitt$, allowing the system to recover between the pulses.
Pulsing was continued until the cell showed signs of failure.
The main experimental finding was that longer current-off periods $t_{\text{off}}$ substantially increased the total stripping capacity~\cite{wu2025}.

\subsection{Two-layer anode model}

Motivated by the above observations, we develop a diffusion-limited model of NaK stripping and seek the protocol that maximizes the stripping capacity.
A minimal model should reproduce the stable-to-unstable stripping transition, its critical current density, the capacity ranges under constant-current stripping, and the observed pulsed-stripping capacities.
However, diffusion through a single finite layer cannot by itself reproduce the stable-to-unstable stripping transition, which suggests the presence of multiple transport scales.

We therefore model Na diffusion through the NaK anode using two layers, comprising a thin transport-limiting layer of length $L_{1}$ and diffusivity $D_{1}$ adjacent to the anode-electrolyte interface, and a bulk NaK anode layer of length $L_{2}$ and diffusivity $D_{2}$
[see Fig.~\ref{fig:diffusion}(c)].
For the transport-limiting layer, we assume $L_{1} \ll L_{2}$ and $D_{1} \ll D_{2}$.
We emphasize that this layer is strictly phenomenological and it should not be interpreted as a distinct material phase.
However, there are several observations that motivate the inclusion of such a layer.
For example, NaK is seen to wet the solid electrolyte poorly, which reduces the true contact area and increases the charge-transfer resistance~\cite{huang2019}.
In addition, impedance measurements further reveal a distinct interfacial contribution associated with phase separation and void formation caused by Na depletion~\cite{wu2025a,shi2025}.

Denoting the molar concentration of Na as $c(x, t)$, the two-layer diffusion equation takes the form
\begin{equation}
  \partial_t c(x, t) = \partial_{x}\left[D(x)\partial_x c(x, t)\right],
  \quad
  0 \leq x \leq L,
  \label{eq:nak_diffusion}
\end{equation}
where $L = L_{1} + L_{2}$ is the total domain length and $D(x)$ is a piecewise-constant diffusivity with values $D_{1}$ and $D_{2}$ in layers 1 and 2, respectively.
The concentration $c(x, t)$ is taken to be initially uniform $c(x, 0) = c_0$ in both layers.
Stripping at a current density of $j(t)$ results in a positive outward flux Faradaic flux $u(t)$ at the domain boundary at $x=0$, while the other end is sealed:
\begin{equation}
  D_{1}\partial_x c(0,t) = u(t) = \left(\frac{1}{\zc F}\right)j(t)
  \quad
  \text{and}
  \quad
  \partial_x c(L, t) = 0.
  \label{eq:nak_bc}
\end{equation}
We also assume perfect contact and inter-layer flux transfer between the two layers~\cite{hickson2009}.
This leads to the following matching conditions that impose continuity of concentration and diffusive flux across the layers:
\begin{equation}
  c(L_{1}^{-},t) = c(L_{1}^{+},t)
  \quad
  \text{and}
  \quad
  D_{1}\partial_x c(L_{1}^{-},t)
  =
  D_{2}\partial_x c(L_{1}^{+},t).
  \label{eq:nak_internal}
\end{equation}
Here, $L_{1}^{-}$ and $L_{1}^{+}$ denote limits taken from the left and right at $x = L_{1}$.

To compare the two-layer model with experiments, we take the initial molar concentration of Na throughout the domain to be $c_{0} = 19.0\unit\text{M}$, corresponding to NaK anodes with a Na mass fraction of 0.5~\cite{wu2025}.
For Na$^{+}$, the charge number $\zc = 1$.
For the bulk anode, we set the diffusivity to%
\footnote{Na diffusivity in NaK ranges from $(4\text{--}6)\times10^{-5}\unitD$ over physically relevant Na concentration range of $0\text{--}40\unit\text{M}$~\cite{gopala-rao1977}.
As this variation is weak we use the mean value for $D_{2}$.}
$D_{2} = 5\times10^{-5}\unitD$
and length to $L_{2} = 0.796\unit\text{mm}$, based on experimental anode volumes.
For the transport-limiting layer, we choose the fitting parameters $L_1=0.05L_2$ and $D_1=2\times10^{-5}D_2$ to match the experimental results.
With these choices, the characteristic diffusive timescales in the two layers are $L_{1}^{2}/D_{1} \approx 4\unit\text{h}$ and $L_{2}^{2}/D_{2} \approx 100\unit\text{s}$, both within the experimentally observed recovery window of $35\unit\text{s}$ to $15\unit\text{h}$.
% Experimentally, recovery is associated with a number of effects, including Na concentration relaxation, NaK realloying, and mechanical recovery~\cite{shi2025}.

\subsection{Constant-current stripping}

To characterize constant-current stripping, we consider two asymptotic regimes corresponding to high and low-to-moderate applied currents.
These limits capture, respectively, rapid interfacial depletion and quasi-steady redistribution of Na throughout the anode.

\paragraph{High stripping currents}
At sufficiently large stripping currents, the Na concentration near the interface is depleted before diffusion has a chance to penetrate through the entire transport-limiting layer.
In this limit, this layer may be treated as semi-infinite and the boundary concentration is given by Sand's result, Eq.~\eqref{eq:inf_transfer},
\begin{equation}
  c(0, t) \approx c_{0} - \frac{2\jc}{\zc F}\left(\frac{t}{\pi D_{1}}\right)^{1/2}.
  \label{eq:nak_high}
\end{equation}

\paragraph{Low stripping currents}
At moderate and low currents, the concentration profile instead evolves quasi-steadily, with the local concentration decreasing approximately at same rate as the spatial average.
To find the quasi-steady concentration profile, we consider the following solution
\begin{equation}
  c(x,t) = \left(c_{0} - \frac{\jc}{\zc FL}t\right) + r(x),
  \label{eq:nak_quasi}
\end{equation}
where the term in parentheses is the mean concentration in the domain, obtained directly from mass conservation, and $r(x)$ is the time-independent deviation from this average.
By definition, $r(x)$ has zero spatial mean.
Upon substituting Eq.~\eqref{eq:nak_quasi} in the diffusion equation, Eq.~\eqref{eq:nak_diffusion}, one finds $[D(x)r'(x)]'=-\jc/(\zc FL)$, with $D_{1}r'(0)=\jc/(\zc F)$ and $r'(L)=0$.
To determine $r(x)$, we impose continuity of concentration and flux at $x=L_{1}$ together with the zero-mean condition.
Substituting the resulting $r(x)$ in Eq.~\eqref{eq:nak_quasi}, and using $L_{1}\ll L_{2}$ and $D_{1}\ll D_{2}$, gives the quasi-steady boundary concentration as
\begin{equation}
  c(0, t) \approx \left(c_{0} - \frac{\jc}{\zc FL}t\right) - \frac{\jc}{\zc F}\left(\frac{L}{3D_{2}} + \frac{L_{1}}{D_{1}}\right).
  \label{eq:nak_low}
\end{equation}

The crossover between the quasi-steady and semi-infinite regimes occurs when the flux due to the applied current becomes comparable to the characteristic diffusive flux ($\sim c_{0}D_{1}/L_{1}$) of the transport-limiting layer.
More explicitly, equating the depletion times obtained from the boundary concentrations in Eqs.~\eqref{eq:nak_high} and \eqref{eq:nak_low} yields the same scaling for the crossover current density $\jc \approx D_{1}c_{0}\zc F/L_{1}$.
Below this current density, diffusion can locally redistribute Na at a faster rate than that at which the mean concentration decreases, leading to a quasi-steady response.
Above it, interfacial Na is depleted faster than it can be replenished from the bulk, producing the strong concentration gradients associated experimentally with unstable stripping.
We therefore identify this crossover with the stable-to-unstable transition seen in the experiments.

For both low and high stripping currents, the usable areal capacity of the anode is the charge transferred per unit area before the boundary concentration vanishes.
Using the depletion times obtained by setting Eqs.~\eqref{eq:nak_high} and \eqref{eq:nak_low} to zero, we see that the stripping capacity for constant currents takes the asymptotic form
\begin{equation}
  \Qc \sim
  \begin{dcases}
    c_{0}\zc FL - \jc L\left(\frac{L}{3D_{2}} + \frac{L_{1}}{D_{1}}\right), & \jc \lesssim D_{1}c_{0}\zc F/L_{1}; \\ \frac{\pi D_{1} c_{0}^{2}\zc^{2}F^{2}}{4\jc}, & \jc \gtrsim D_{1}c_{0}\zc F/L_{1}.
  \end{dcases}
  \label{eq:nak_asymptotic}
\end{equation}
As $\jc\to 0$, the stripping capacity approaches the theoretical maximum $\Qc\to c_{0}\zc FL$, which is approximately $40.6\unitQ$ for the NaK anodes of Ref.~\cite{wu2025}.
The predicted stable-to-unstable transition at $\jc\approx 0.46\unitj$ is in reasonable agreement with the experimental transition near $0.6\unitj$.
At a stable stripping current density of $\jc=0.25\unitj$, Eq.~\eqref{eq:nak_asymptotic} predicts $\Qc \approx 18.6\unitQ$, in good agreement with the experimentally observed capacity of $20\unitQ$.
Likewise, at an unstable current density of $\jc = 1\unitj$, Eq.~\eqref{eq:nak_asymptotic} predicts a capacity of about $\Qc \approx 0.73\unitQ$, which is again consistent with observed capacities of less than $1\unitQ$~\cite{wu2025}.

\begin{figure}{\centering\includegraphics{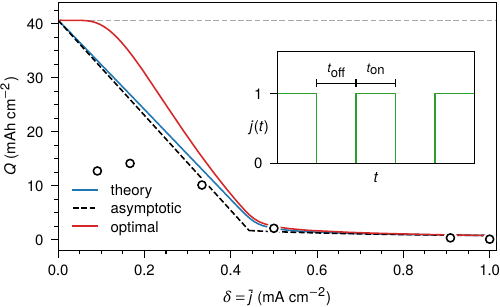}}
  \caption{
    Areal capacity $Q$ as a function of the duty cycle $\delta = t_{\text{on}}/(t_{\text{on}} + t_{\text{off}})$ during pulsed stripping.
    The current-on duration is fixed at $t_{\mathrm{on}} = 1\unit\text{s}$, while the current-off duration $t_{\text{off}}$ is varied, with a maximum current density of $j_{0} = 1\unitj$.
    The inset shows the pulsed current density profile $j(t)$.
    For these pulses, the numerical value of the duty cycle $\delta$ equals the average current density $\bar{j} = j_{0}\delta$.
    The theoretical capacities (solid blue curves) were obtained using a double-layer anode model, and the corresponding asymptotic results (dashed black curves) using Eq.~\eqref{eq:nak_asymptotic}.
    Both use parameters relevant to the experimental NaK anodes of Ref.~\cite{wu2025}, which have a maximum areal capacity of about $40.6\unitQ$.
    Open circles denote the experimental capacities for pulsed-current protocols.
    The optimal capacities for the continuous bang--ride protocol are shown by the solid red curve, using the same total duration as the corresponding pulsed-current protocols.
    The elbow in the theoretical curves separates the stable and unstable stripping regimes and occurs at $\bar{j} \approx c_0 D_1 F/L_1 \approx 0.46\unitj$, in reasonable agreement with the experimentally observed transition at approximately $\bar{j} \approx 0.6\unitj$~\cite{wu2025}.
  }
  \label{fig:nak}
\end{figure}

\subsection{Stripping under pulsed and optimal currents}

The two-layer diffusion model also admits a transfer-function representation relating the boundary concentration to the applied current density $j(t)$, as in Eqs.~\eqref{eq:inf_transfer} and \eqref{eq:finite_transfer}.
The exact transfer function, written as a modal expansion, is derived in Appendix~\ref{app:transfer}.
We use it to study the same pulsed-stripping protocol as in the experiments, with a maximum current density of $j_{0}=1\unitj$, a fixed on-time $t_{\mathrm{on}}=1\unit\text{s}$, and variable off-time $t_{\mathrm{off}}$.
Figure~\ref{fig:nak} shows the resulting areal capacity $Q$ as a function of the duty cycle $\delta=t_{\mathrm{on}}/(t_{\mathrm{on}}+t_{\mathrm{off}})$.
Also shown are the asymptotic capacities from Eq.~\eqref{eq:nak_asymptotic}, evaluated using the time-averaged current $\bar{j}=\delta j_{0}$, which closely approximate the transfer-function results.
Furthermore, the figure clearly distinguishes the stable and unstable stripping regimes, with the elbows in the capacity curves marking the transition between the two.

Pulsed-stripping experiments can take several hours to several days as the duty cycle decreases, and only six capacity--duty-cycle measurements have been reported in Ref.~\cite{wu2025}.
As we see from Fig.~\ref{fig:nak}, for duty cycles above $0.2$, the experimental capacities agree well with the two-layer model, whereas the capacities for the two lowest duty cycles deviate substantially from the predictions.
These deviations may reflect degradation of the NaK anode (unrelated to stripping) during the much longer experiments.
Moreover, the low capacities at duty cycles below $0.2$ appear inconsistent with the stable stripping observed at a constant current density of $0.25\unitj$~\cite{wu2025}, further suggesting that NaK degradation due to long experimental durations may have contributed to the lowering of the capacities.

The current protocol that maximizes the transferred charge subject to the boundary concentration constraint retains the bang--ride form.
The riding current is most easily obtained by numerically solving~\cite{trefethen2000} the diffusion equation with homogeneous Dirichlet--Neumann boundary conditions, as described in Sec.~\ref{sec:charging}.
Figure~\ref{fig:nak} shows the resulting optimal areal capacity for the two-layer NaK anode model assuming a boundary concentration constraint $c(0,t)\geq 0$.
The optimal bang--ride protocol consistently outperforms pulsed stripping because no charge is transferred during the ``off'' periods of a pulse protocol, whereas the optimal protocol reduces the current only as required to satisfy the concentration constraint while continuing to transfer charge.

We can also compare the optimal protocol with the best constant-current protocol, as in the previous sections.
Appreciable gains from bang--ride control occur only when the maximum allowable current is large.
In this limit, the two-layer model approaches the semi-infinite regime, for which the maximum capacity gain is approximately $27\%$, as in the Li electrodeposition experiments discussed in Sec.~\ref{sec:electrodeposition}.

\section{Conclusions}
\label{sec:conclusion}

In many electrochemical systems, diffusion cannot replenish or remove material at an interface as rapidly as the current demands.
Motivated by this limitation, we studied optimal current-control protocols for 1D diffusion subject to a constraint on the interfacial concentration.
Using the exact transfer-function formulation of the diffusion equation, we showed that the optimal current has a bang--ride form.
The maximum admissible current is used until the concentration constraint becomes active, after which the current is progressively reduced to maintain that constraint.

For diffusion in a semi-infinite half-line, the optimal current can be obtained analytically.
This makes it possible to compare the optimal protocol directly with the best constant-current protocol over the same time interval.
With capacity defined as the amount of charge transferred over a given time interval, the optimal protocol yields a maximum capacity gain of about $27\%$ over the best constant-current protocol, independent of the material parameters.
This gives a sharp limit on the improvement that can be obtained from optimal current control in the homogeneous semi-infinite problem.
For the corresponding fast-charging problem, the same analysis gives a maximum reduction of about 38\% to the charging time, relative to the best constant-current protocol.
We also compared continuous optimal currents with pulsed-current protocols.
During each off period, the boundary concentration recovers but no charge is transferred.
As a result, pulsed protocols transfer less charge over the same time interval, and the continuous bang--ride protocol consistently performs better.

Finally, we extended the formulation to finite multilayer diffusion, where similar diffusion limitations apply.
We used this framework to model NaK liquid-anode experiments with a two-layer diffusion system and determine discharge protocols that increase charge transfer while maintaining the concentration constraint.
The same bang--ride formulation applies to both lower and upper concentration limits and to both charging and discharging problems.

More broadly, our analysis shows that diffusion-limited transport with a constrained boundary concentration can be treated as a control problem, independent of whether the boundary is depleted or saturated.
The resulting optimal protocols and performance bounds distinguishes improvements achievable through control alone, without changing the geometry or transport properties.
Natural next steps might in fact include these effects.
As noted earlier, drying in inhomogeneous porous materials~\cite{pel2002,varadharaju2001} provides one example where strong moisture gradients can lead to case hardening or cracking~\cite{tsapis2005,gulati2015}.
Similar multilayer diffusion problems arise in transdermal drug delivery, where the delivery rate decreases as the drug near the skin is depleted~\cite{defraeye2020}.
Extensions to our theory, such as allowing for changes in geometry or transport properties, could further help optimize the drying and release protocols in these examples.

\bigskip

\paragraph*{Acknowledgments}
We thank Paul Braun, Kelsey Hatzell, Yannis Kevrekidis, Matthew McDowell, Jungki Min, Partha Mukherjee, Aditya Singla, Sumit Sinha, Sun Geun Yoon, Beniamin Zahirisabzevar, and other members of the DARPA MINT collaboration for useful conversations.
This work was supported by the DARPA MINT program, the Simons Foundation, and the Henri Seydoux Fund.

\paragraph*{Data availability}
All our numerical codes as well as the data used for experimental comparison is available publicly~\cite{github}.

\appendix

\section{Optimality of the bang--ride protocol for diffusion-limited systems}
\label{app:optimality}

To verify the optimality of the bang--ride protocol in Eq.~\eqref{eq:bang_ride}, we follow classical optimal-control approaches for systems governed by Volterra integral equations~\cite{vinokurov1969,kamien1976}.
For expository convenience, we take the Faradaic flux $u(t)$, rather than the current density $j(t)$, as the control variable---in all the cases we consider in this paper, the two differ only by a multiplicative constant, see, e.g., Eq.~\eqref{eq:inf_bc}.
The objective is to maximize the total amount of ``material'' transferred,
\begin{equation}
  \mathcal{C}[u(t)] = \int_{0}^{T}\dd{t}\, u(t).
\end{equation}
subject to the boundary-concentration constraint $c(0,t) - \chi c_{0} \geq 0$ and the flux constraints $0 \leq u(t) \leq u_{0}$.
Let $p(t)\geq 0$ denote the Lagrange multiplier associated with the boundary-concentration constraint.
Writing the boundary concentration $c(0, t)$ in terms of the transfer function $Z(t)$, as in Eq.~\eqref{eq:inf_transfer}, we then construct the auxiliary functional
\begin{equation}
  \mathcal{J}[u(t),p(t)] = \int_{0}^{T}\dd{t}\,\left\{u(t) + p(t)\left[\left(1 - \chi\right)c_{0} - \int_{0}^{t}\dd{s}\,Z(t - s)u(s)\right]\right\}.
\end{equation}
As $p(t) \geq 0$, we see that $\mathcal{J} \geq \mathcal{C}$ always.
The transfer function $Z(t)$ is assumed to be positive and strictly decreasing.
In order to examine variations of $\mathcal{J}$ with respect to $u(t)$, we change the inner integral's order of integration to find
\begin{equation}
  \mathcal{J}[u(t),p(t)] = \int_{0}^{T}\dd{t}\,\left[\Psi_{p}(t)\,u(t) + \left(1 - \chi\right)c_{0}p(t)\right],
\end{equation}
where $\Psi_{p}$ is a switching function defined through a backward integral involving the multiplier $p$ and given by
\begin{equation}
  \Psi_{p}(t) = 1 - \int_{t}^{T}\dd{s}\, Z(s - t)\,p(s).
\end{equation}
Because $\mathcal{J}$ is linear in $u$, its maximum over $0\leq u(t)\leq u_0$ is determined by the sign of $\Psi_{p}(t)$.
More specifically, any $u(t)$ that maximizes $\mathcal{J}$ must be of the form~\cite{kirk1970}
\begin{equation}
  u(t) =
  \begin{cases}
    u_{0},                        & \Psi_{p}(t) > 0; \\
    \text{any } u \in [0, u_{0}], & \Psi_{p}(t) = 0; \\
    0,                            & \Psi_{p}(t) < 0.
  \end{cases}
  \label{app:eq:switches}
\end{equation}

Our goal is to construct a multiplier $\ps \geq 0$ that is consistent with the above switching condition for the protocol in Eq.~\eqref{eq:bang_ride}.
Consider,
\begin{equation}
  \ps(t) =
  \begin{cases}
    0, & t < \ts;\\
    p(t) \geq 0 \text{  such that  } \Psi_{p}(t) = 0, & t \geq \ts.
  \end{cases}
  \label{app:eq:nakplier}
\end{equation}
For a general transfer function, we assume that the backward Volterra equation $\Psi_{p}(t)=0$ in the riding phase ($t \geq \ts$) admits a solution $p(t) \geq 0$.
As an example, for the semi-infinite transfer function $Z(t) = (\pi D t)^{-1/2}$, it can be directly checked that the positive multiplier $p(t) = D^{1/2}[\pi(T - t)]^{-1/2}$ satisfies $\Psi(t) = 0$ for $t \geq \ts$.

For $t < \ts$, using $p=\ps$, we have
\begin{equation}
  \Psi_{\ps}(t) = 1 - \int_{\ts}^{T}\dd{s}\, Z(s-t)\,\ps(s)
  >
  1 - \int_{\ts}^{T}\dd{s}\, Z(s-\ts)\,\ps(s) = 0.
\end{equation}
Here, we have used $s-t > s-\ts$, the strict decrease of $Z$, and the fact that $\ps\geq0$.
Hence, $\Psi_{\ps}(t) > 0$ before $\ts$, whereas $\Psi_{\ps}(t)=0$ during the riding phase.
The former forces the optimal flux $\us$ to be equal to $u_{0}$ for $t < \ts$, while the latter permits any allowable flux $0 \leq u \leq u_{0}$, including the flux corresponding to the riding current $\jr$.
It follows that for the multiplier $\ps$ in Eq.~\eqref{app:eq:nakplier}, the flux $\us$ corresponding to the current in Eq.~\eqref{eq:bang_ride} maximizes $\mathcal{J}$.

It remains to show that $\us$ also maximizes the original objective functional $\mathcal{C}$.
For any feasible flux $u$, the constraint $c(0, t) - \chi c_{0} \geq 0$, and the multiplier $\ps(t)\geq0$ imply
\begin{equation}
  \mathcal{C}[u]
  \leq \mathcal{J}[u, \ps]
  \leq \mathcal{J}[\us, \ps]
  = \mathcal{C}[\us].
\end{equation}
For $u = \us$, the multiplier term in $\mathcal{J}$ vanishes because either $\ps(t)=0$ or $c(0, t) = \chi c_{0}$, which gives the last equality.
Therefore, $\mathcal{C}[u]\leq\mathcal{C}[\us]$ for every feasible flux $u$, proving that the bang--ride protocol $\us$ is optimal.%
\footnote{This is essentially a primal--dual argument from convex optimization~\cite{boyd2004}: the constructed multiplier makes $\us$ maximize the $\mathcal{J}$, while the multiplier term vanishes along $\us$.
  The resulting upper bound on $\mathcal{C}$ is therefore equal to $\mathcal{C}[\us]$.
}

\section{Transfer functions}
\label{app:transfer}

\subsection{Finite-domain diffusion}

To find the boundary concentration associated with the finite-layer diffusion equation, Eq.~\eqref{eq:finite_diffusion}, we consider a solution of the form
\begin{equation}
  c(x, t) = b(x, t) + \frac{u(t)}{2DL}x^{2} - \frac{u(t)}{D}x,
  \label{eq:finite_solution}
\end{equation}
where $b(x, t)$ is the transient solution and the sole purpose of the remaining terms is to satisfy the boundary conditions in Eq.~\eqref{eq:finite_bc}.
Substituting Eq.~\eqref{eq:finite_solution} in Eq.~\eqref{eq:finite_diffusion}, we that the transient solution $b(x, t)$ satisfies a diffusion equation of the form
\begin{subequations}
  \begin{align}
    \partial_{t}b(x, t) &= D\partial_{x}^{2}b(x, t) + a(x, t),\\
    \label{eq:finite_transient}
    b(x, 0) &= -\frac{u(0)}{2DL}x^{2} + \frac{u(0)}{D}x,\\
    \partial_{x}b(0, t) &= \partial_{x}b(L, t) = 0.
  \end{align}
\end{subequations}
In Eq.~\eqref{eq:finite_transient}, the source term $a(x, t)$ depends on the current $u(t)$ and its time derivative $u'(t)$, and is given by
\begin{equation}
  a(x, t) = \frac{u(t)}{L} - \frac{u'(t)}{2DL}x^{2} + \frac{u'(t)}{D}x.
\end{equation}
Equation~\eqref{eq:finite_transient} is in the form of a standard initial boundary value diffusion problem with a source term.
The solution can be formally written down using Duhamel's principle as
\begin{equation}
  b(x, t) = \int_{0}^{L}\dd{x'}\,G(x, x', t)\, b(x', 0) + \int_{0}^{t}\dd{s}\,\int_{0}^{L}\dd{x'}\,G(x, x', t-s)\,a(x',s),
  \label{eq:trans_single}
\end{equation}
where the $G(x, x', t)$ is the finite-domain diffusion kernel given by~\cite{carslaw1959}
\begin{equation}
  G(x, x', t) = \frac{1}{L} + \frac{2}{L}\sum_{m = 1}^{\infty}\cos\left(\frac{\pi m x}{L}\right)\cos\left(\frac{\pi m x'}{L}\right)\exp\left(-\frac{\pi^{2}m^{2}Dt}{L^{2}}\right).
\end{equation}
In Eq.~\eqref{eq:trans_single}, integrating the terms involving $u'(t)$ by parts cancels all contributions from the first integral.
The boundary concentration $c(0, t) = b(0, t)$ in Eq.~\eqref{eq:finite_transfer} is then obtained after straightforward algebra.

\subsection{Multilayer domains}

For layered diffusion, the transfer function can be derived in the Laplace domain using a transfer-matrix formulation~\cite{carslaw1959,vliet1980}.
It can be also obtained from the weak formulation of the diffusion equation through an eigenmode expansion as we show below.
Consider the following time-independent eigenvalue problem with homogeneous Neumann boundary conditions:
\begin{equation}
  \partial_{x}\left[D(x)\partial_{x}\phi_{m}(x)\right] = -\lambda_{n}\phi_{m}(x),
  \quad \partial_{x}\phi_{m}(0) = \partial_{x}\phi_{m}(L) = 0,
  \label{app:eq:eigenvalue}
\end{equation}
Here, $D(x)$ is the piecewise constant diffusivity appearing in Eq.~\eqref{eq:nak_diffusion}, and the $m$th eigenvalue and normalized eigenmode are $\lambda_{m}$ and $\phi_{m}$, respectively.
It can be shown that the diffusion operator in Eq.~\eqref{app:eq:eigenvalue} with the matching conditions, Eq.~\eqref{eq:nak_internal}, is self-adjoint~\cite{hickson2009}, making $\lambda_{m}$ real and $\phi_{m}$ mutually orthogonal.

A modal expansion of the concentration
\begin{equation}
  c(x, t) = c_{0} + \sum_{m = 0}^{\infty} a_{m}(t)\phi_{m}(x),
  \label{app:eq:modal}
\end{equation}
in terms of the modes in Eq.~\eqref{app:eq:eigenvalue} will \emph{not} satisfy the time-dependent two-layer diffusion equation, Eq.~\eqref{eq:nak_diffusion}, owing to the inhomogeneity in the boundary conditions.
However, such a modal expansion may be used in the weak form of the diffusion equation to find the modal coefficients $a_{n}(t)$~\cite{haberman1987,reddy1993}.

To construct the weak form of Eq.~\eqref{eq:nak_diffusion}, we multiply it by a test function $\phi$ that is twice-differentiable in $[0, L]$ and integrate by parts once, which results in
\begin{equation}
  \int_{0}^{L}\dd{x}\, \partial_{t}c(x,t)\,\phi(x) = -u(t)\phi(0) - \int_{0}^{L}\dd{x}\,D(x)\partial_{x}c(x, t)\phi'(x),
  \label{app:eq:weak}
\end{equation}
where we have also made use of the boundary conditions in Eq.~\eqref{eq:nak_bc}.
Using $\phi = \phi_{m}(x)$ and Eq.~\eqref{app:eq:modal} in Eq.~\eqref{app:eq:weak} and integrating by parts once again, we see that the modal coefficients $a_{m}(t)$ satisfy
\begin{equation}
  \partial_{t}a_{m}(t) + \lambda_{m}a_{m}(t) = -u(t)\phi_{m}(0),
\end{equation}
with the Faradaic flux $u(t)$ acting as a forcing term.
For a uniform initial concentration, we have $a_{m}(0) = 0$.
Hence, the above equation has the solution
\begin{equation}
  a_{m}(t) = -\phi_{m}(0)\int_{0}^{t}\dd{s}\,\e^{-\lambda_{m}(t - s)}u(s),
\end{equation}
This gives the weak solution of Eq.~\eqref{eq:nak_diffusion} as
\begin{equation}
  c(x, t) = c_{0} - \int_{0}^{t}\dd{s}\,\left[\sum_{m=0}^{\infty} \phi_{m}(0)\phi_{m}(x)\,\e^{-\lambda_{m}(t-s)}\right]u(s).
  \label{eq:nak_solution}
\end{equation}

Setting $x = 0$ in Eq.~\eqref{eq:nak_solution} yields the boundary concentration as
\begin{equation}
  c(0, t) = c_{0} - \int_{0}^{t}\dd{s}\,Z(t - s)\,u(s),
  \quad
  \text{with}
  \quad
  Z(t) = \sum_{m=0}^{\infty} \abs{\phi_{m}(0)}^{2}\,\e^{-\lambda_{m}t}.
  \label{eq:nak_transfer}
\end{equation}
We see that the boundary concentration for the multilayer problem is in the same form as Eq.~\eqref{eq:inf_transfer} with a positive, strictly decreasing transfer function $Z(t)$.
For the double-layer NaK anode model, Eq.~\eqref{eq:nak_transfer} is used to find the boundary concentration for pulsed stripping, from which the anode capacity can be estimated.

\bibliography{library,misc}

\end{document}